\UseRawInputEncoding
\documentclass[%
 aps,
 prd,
 twocolumn,
 groupedaddress,  
 nofootinbib,
 amsmath,amssymb,
 floatfix
]{revtex4-2}
\usepackage{natbib}
\usepackage{graphicx}
\usepackage{adjustbox}
\usepackage{dcolumn}
\usepackage{bm}
\usepackage[dvipsnames]{xcolor}
\usepackage{color, colortbl}
\usepackage{relsize}
\usepackage[mathlines]{lineno}

\usepackage{stackengine}
\stackMath
\usepackage{xcolor}
\usepackage{dcolumn}
\usepackage{bm}
\usepackage{physics}  
\usepackage{comment}
\usepackage{hyperref}
\usepackage{float}
\usepackage{soul}

\begin{document}

\title{Gravitational waves from primordial black holes passing by neutron stars: observational prospects for the Galactic center}

\author{Nicolas Esser$^1$}
\email[]{nicolas.esser@ulb.be} 

\author{Juan Garc\'ia-Bellido$^2$}
\email[]{juan.garciabellido@uam.es}

\author{Peter Tinyakov$^1$}
\email[]{petr.tiniakov@ulb.be}

\affiliation{$^1$ Service de Physique Th\'eorique, Universit\'e libre de Bruxelles, Boulevard du Triomphe, CP225, 1050 Brussels, Belgium}

\affiliation{$^2$ Instituto de F\'isica Te\'orica UAM/CSIC,  Universidad Aut\'onoma de Madrid, Cantoblanco 28049 Madrid, Spain}

\date{\today}

\begin{abstract}

We investigate the gravitational wave (GW) signals emitted by planetary-mass primordial black holes (PBHs) passing nearby or traversing neutron stars (NSs). While previous studies mainly focused on the detailed waveforms of the signals, we estimate the rate of PBH-NS gravitational-wave events originating from the Galactic center and compute the probability of detecting a signal over 10 years of LIGO-Virgo-KAGRA observations. We examine in detail the case of PBHs bound to NSs, focusing on eccentric orbits that give rise to repeated GW bursts emitted in correlated series, each burst corresponding to a periastron passage. Despite the enhancement from the large number of bursts produced by a single PBH-NS pair, the total number of signals produced in this way remains subdominant to those due to random unbound encounters of PBHs with NSs. We also find that both types of signals have a very small probability $P\lesssim 10^{-8}$ to be detected in a $10$ year period. 

\end{abstract}

\preprint{IFT-UAM/CSIC-25-169,
ULB-TH/26-05}

\maketitle

\section{Introduction}

Primordial black holes (PBHs), hypothesized more than five decades ago \cite{Zeldovich,Hawking} as objects that could have formed in the early Universe, are now widely discussed as a compelling dark matter (DM) candidate~\cite{Carr:2023tpt}. Their existence would have significant cosmological and astrophysical implications -- for example, as a fraction or even the entirety of DM; as seeds for the formation of supermassive black holes \cite{Clesse:2015wea}; as catalysts for baryon number generation \cite{Garcia-Bellido:2019tvz}; and in other roles. The lack of observational evidence has placed strong constraints on their abundance \cite{Carr_review,Green_2020}; however, these constraints do not completely rule out their existence~\cite{Carr:2023tpt}.

A promising way to probe PBHs is gravitational-wave (GW) astronomy. In particular, mergers of binary primordial black holes of stellar masses have been extensively investigated from both theoretical (see e.g. \cite{Raidal_2024}) and observational (see e.g. \cite{Garcia-Bellido:2017fdg}) perspectives, as they would generate signals within the sensitivity range of current GW detectors such as LIGO, Virgo, and KAGRA (LVK) \cite{Clesse:2016vqa}.

In this paper, we explore an alternative possibility for detecting primordial black holes through gravitational-wave signals. We focus on burstlike signals produced when PBHs pass near Galactic neutron stars (NSs). Two main reasons make this approach potentially promising. First, neutron stars are almost as compact as stellar-mass black holes  --  an essential factor for producing strong GW signals  --  while being significantly more abundant. Second, the characteristic GW frequency is set by the orbital dynamics, which depend on the total mass of the PBH-NS system. For light PBHs, the total mass is approximately equal to the NS mass, making it comparable to that of an NS-NS binary. The resulting GW frequency can be estimated as
\begin{equation}
f_{\rm GW} \simeq 100~{\rm Hz} \left(\frac{M}{M_\odot}\right)^{1/2}
\left(\frac{100~{\rm km}}{a}\right)^{3/2},
\label{eq:freq_GW}
\end{equation}
where $M$ is the neutron star mass, which is equivalent to the total mass in the case of a light PBH, and $a$ is the characteristic orbital scale. Note that for highly eccentric orbits relevant in what follows, $a$ is determined by the periastron distance. As is seen from Eq.~\eqref{eq:freq_GW}, the resulting GW frequencies naturally fall within the LVK sensitivity range, around $\mathcal{O}(100-1000)$ Hz, even for PBH masses $m\ll M_\odot$, thereby allowing one to probe this otherwise hardly accessible mass range. 

Gravitational wave emission from PBH-NS systems has previously been studied in the context of PBH capture by neutron stars \cite{Genolini_2020,Holst_2025}. These works focused on the energy loss due to gravitational radiation and dynamical friction from the NS material, and how these processes could ultimately lead to PBHs being captured in neutron stars. Captured PBHs would emit a continuous GW signal as they spiral inward \cite{Gao_2023,Baumgarte_2024}, and would transmute their hosts into stellar-mass black holes within a short time \cite{Capela_2012}. However, since PBH abundance is constrained from above by that of dark matter, the capture rate of PBHs by NSs in the Milky Way appears to be too low to allow realistic detection (see e.g. \cite{Tinyakov_2024}).

What has not been considered, to the best of our knowledge, is the detection of signals at an earlier stage, when bound PBHs are orbiting {\em outside} the NS and generating burstlike signals at each periastron passage. The repeated nature of these passages could increase the event rate and thus the likelihood of detection, potentially making the observation of such events a realistic possibility. The purpose of the present work is to investigate this scenario, and assess what are the most likely physical configurations that may lead to detectable GW signals from PBH-NS pairs. 

The first question is which PBH mass range is most promising for detection. The absolute signal strength increases linearly with the PBH mass $m$, and so does the detection horizon, reaching hundreds of Mpc for $m$ approaching $M_\odot$. The rate of detectable events therefore depends on the PBH mass through two competing effects: an increase in the number of neutron stars enclosed within the detection horizon, and a decrease in the PBH number density proportional to $1/m$. Because the Galactic center contains a large density of both neutron stars and dark matter, the rate will exhibit a local maximum when the detection horizon barely reaches it. Given its distance of $d=8.2$~kpc, this corresponds to PBH masses $m \sim 10^{-4} M_\odot$ on which we will focus in what follows.

The key question, however, is the absolute value of the event rate: do such PBH-NS encounters happen often enough to be detected in a reasonably near future? 
For hyperbolic, unbound encounters the estimate is straightforward: \begin{align}
\begin{split}
    \Gamma&\sim N_\text{NS}  n_\text{PBH}\times\pi r_\text{det}^2\times \frac{2GM}{r_\text{det}\bar{v}^2}\times \bar{v}\\[3pt]
    &\sim 10^{-9}\text{ yr}^{-1}
    \left(\frac{N_\text{NS}}{10^9}\right)\left(\frac{f_\text{PBH}}{1}\right)\left(\frac{\rho_\text{DM}}{10\text{ GeV/cm}^3}\right)\\[5pt]&\hspace{1.27cm}\times\left(\frac{10^{-4}M_\odot}{m}\right)\left(\frac{r_\text{det}}{100\text{ km}}\right)\left(\frac{200\text{ km/s}}{\bar{v}}\right),
\label{eq:rate_unbound}
\end{split}
\end{align}
where $N_\text{NS}$ is the number of neutron stars in the considered region (in our case, the Milky Way), $r_\text{det}$ denotes the largest periastron distance at which an encounter in the Galactic center still produces a detectable signal on Earth, $\bar{v}$ is the typical relative velocity of PBHs and NSs, and $n_\text{PBH}=f_\text{PBH}\rho_\text{DM}/m$ is the PBH number density, with $f_\text{PBH}$ the fraction of DM in the form of PBHs and $\rho_\text{DM}$ the dark matter density. The third factor accounts for gravitational focusing \cite{BinneyTremaine}. We have checked that this estimate is valid for any smooth and compact relative velocity distribution with a characteristic velocity $\bar v$, up to order-one corrections. For instance, in the case of a Maxwellian distribution the difference with Eq.~\eqref{eq:rate_unbound} would be of order $\sim40\%$. 

The rate of direct encounters (\ref{eq:rate_unbound}) is, obviously, too low. 
However, when PBHs are trapped on bound eccentric orbits around neutron stars, they {\em repeatedly} pass close to the NS, emitting GW signals at each periastron passage. Estimating the contribution of such bound PBHs to the detectable event rate is less straightforward, since the encounters occur as series: initially rare, but becoming increasingly frequent as the orbit decays toward the final inspiral. Over time, different series -- originating from different PBHs -- occur randomly and independently, whereas the encounters within a given series are statistically correlated. In this context, the relevant question is: given an observation time window of, say,  $t_\text{obs}\sim 10$~yr, what is the probability of detecting at least one such encounter during this period?

To address this question, we divide the calculation into several parts. We begin in Sec.~\ref{sec:strain} by characterizing the expected GW signals from PBH-NS encounters, computing the associated strain and signal-to-noise ratio to determine which PBH orbits yield detectable signals. We then evaluate the signal rate in multiple steps. First, in Sec.~\ref{sec:boundPBHs}, we calculate the rate at which new bound PBH-NS systems form in the Milky Way. Next, in Sec.~\ref{sec:furtherevoPBH}, we evaluate the energy lost by PBHs during their encounters with neutron stars, through both gravitational-wave emission and dynamical friction, the latter being the dominant mechanism when the PBH follows a trajectory that crosses the neutron star. We compute the resulting orbital evolution of PBHs and also derive a condition on the maximum allowed apastron for which PBHs remain bound to neutron stars despite perturbations from neighboring stars. Finally, in Sec.~\ref{sec:probasignal}, we compute the probability of detecting a signal from PBHs orbiting neutron stars in the Galactic center over a 10-year period, assuming the current sensitivity of LVK. We present our final results and conclusions in Sec.~\ref{sec:conclusion}.

We find that the probability to detect a signal is small, and that the most likely detection scenario remains that of unbound PBHs. Although bound PBHs undergo repeated passages, this repetition does not fully compensate for their lower number density compared to the unbound population. Even the presence of a strong dark matter spike in the Galactic center, and improved sensitivity -- such as that expected from the future Einstein Telescope -- cannot provide a reasonable detection probability. 

\section{GW signal and strain}
\label{sec:strain}
The gravitational waves emitted by a PBH passing close to a neutron star on an unbound or eccentric orbit have a characteristic GW strain that can be expressed as a function of $P(f)$, the power radiated at frequency $f$, as~\cite{JGB_2018}
\begin{equation}
h_c(f)\simeq\frac{1}{d}\sqrt{\frac{4G}{c^3}P(f)}    
\label{eq:strain_from_power}
\end{equation}
where $d$ is the distance to the observer. The power $P(f)$ has been calculated for hyperbolic orbits in \cite{JGB_2018}. For eccentric elliptic orbits which are of interest to us here, the power was derived in \cite{Berry_2010}. They also derived the power in the parabolic limit of eccentricity $e=1$ (see their Eq.~(23)). For simplicity, and because most of the PBH bound orbits contributing to the event rate are very eccentric, we use that expression as an estimate. It reads
\begin{equation}
    P(f)=\frac{4\pi^2}{5}\frac{G^3}{c^5}\frac{m^2M^2}{r_p^2}l\left(\frac{f}{f_c}\right),
\label{eq:power}
\end{equation}
where $f_c=\sqrt{GM/r_p^3}/(2\pi)$ is the  characteristic frequency equal to the frequency of a circular Newtonian orbit of size $r_p$, with $r_p$ the periastron distance. The function $l(f/f_c)$ is a dimensionless factor described in \cite{Berry_2010} that combines multiple Bessel functions.

\begin{figure}[t!]
    \centering
    \includegraphics[width=\linewidth]{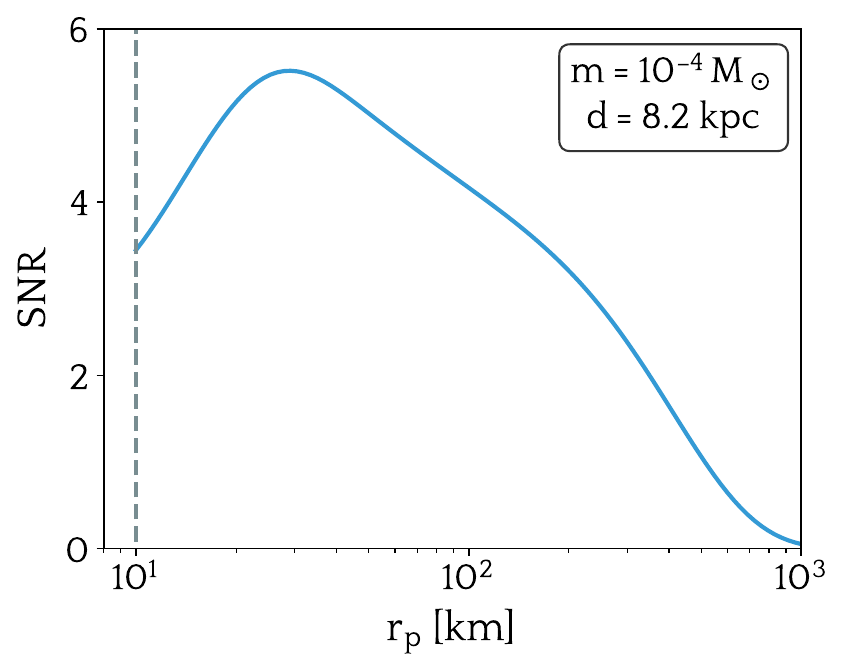}
    \caption{SNR of the GW signal in LVK O4 as a function of $r_p$ for a PBH orbiting a neutron star in an elliptic or hyperbolic orbit with eccentricity close to $1$. The neutron star radius $R=10$ km is displayed as a dashed vertical line. The PBH mass is fixed to $m=10^{-4}M_\odot$ and the distance of the PBH-NS system with respect to the Earth is $d=8.2$ kpc. The SNR scales as $\propto m/d$, so the result can be straightforwardly rescaled to any PBH mass and distance to the Earth.}
    \label{fig:SNR}
\end{figure}

The characteristic strain is related to the Fourier transform of the physical strain
$\tilde h(f)$
via $h_c^2(f)=4f^2\tilde h^2(f)$ \cite{Moore_2014}.
Given the strain $\tilde h(f)$ and the Power Spectral Distribution (PSD) $S_n(f)$ associated with the noise in the GW detector, the signal-to-noise ratio can be computed as \cite{Maggiore}:
\begin{equation}    \text{SNR}=2\left[\int_{f_\text{min}}^{f_\text{max}}\frac{\tilde h^2(f)}{S_n(f)}df\right]^{1/2},
    \label{eq:SNR}
\end{equation}
where $f_\text{min}$ and $f_\text{max}$ are the minimum and maximum gravitational wave frequencies in the sensitivity range of the detector. In practice, for an order of magnitude computation of the SNR, one can approximate the PSD by a square well with depth $S_0$ and frequency limits $[f_{\rm min},f_{\rm max}]$. In this approximation, the SNR is determined by the integral of $\tilde h^2(f)$ between $f_{\rm min}$ and $f_{\rm max}$. The LVK sensitivity from run O4 has amplitude spectral density $S_0^{1/2} = 4\times10^{-24}/\sqrt{\rm Hz}$, $f_{\rm min}=10$ Hz and $f_{\rm max}=800$ Hz~\cite{LIGOScientific:2025hdt}. Using these values, one can compute the SNR for a single encounter event, which only depends on the periastron distance of the PBH with respect to the NS, and on the masses of the two objects. In Fig.~\ref{fig:SNR}, we show the SNR as a function of $r_p$ for a PBH of mass $m=10^{-4}M_\odot$ passing close to a neutron star in the Galactic center.

From Fig.~\ref{fig:SNR}, and because the SNR scales linearly with $m$ (see Eqs.~\eqref{eq:strain_from_power}, \eqref{eq:power} and \eqref{eq:SNR}), it is clear that gravitational waves arising from the Galactic center may only be detected for PBH masses $m\gtrsim 10^{-4}\,M_\odot$. Note that the above calculations rely on the point-mass approximation, which is no longer valid once the PBH orbit crosses the neutron star at $r_p\lesssim R=10$~km. Importantly, the SNR is \textit{not} maximized at $r_p=R$. This is because for passages too close to the NS, with $r_p\lesssim 100\,\mathrm{km}$, a significant fraction of the emitted energy is carried by gravitational waves with frequencies that are too high compared to the LVK sensitivity band, which consequently reduces the SNR for these instruments. For PBHs crossing neutron stars, we will assume for simplicity that the SNR is independent of $r_p$ and equal to its value at \( r_p = R \). For detailed calculations of the gravitational waves emitted by PBHs inside neutron stars, see e.g. Refs.~\cite{Baumgarte_2024,Baumgarte:2024_2}.

\begin{figure}[t!]
    \centering
    \includegraphics[width=\linewidth]{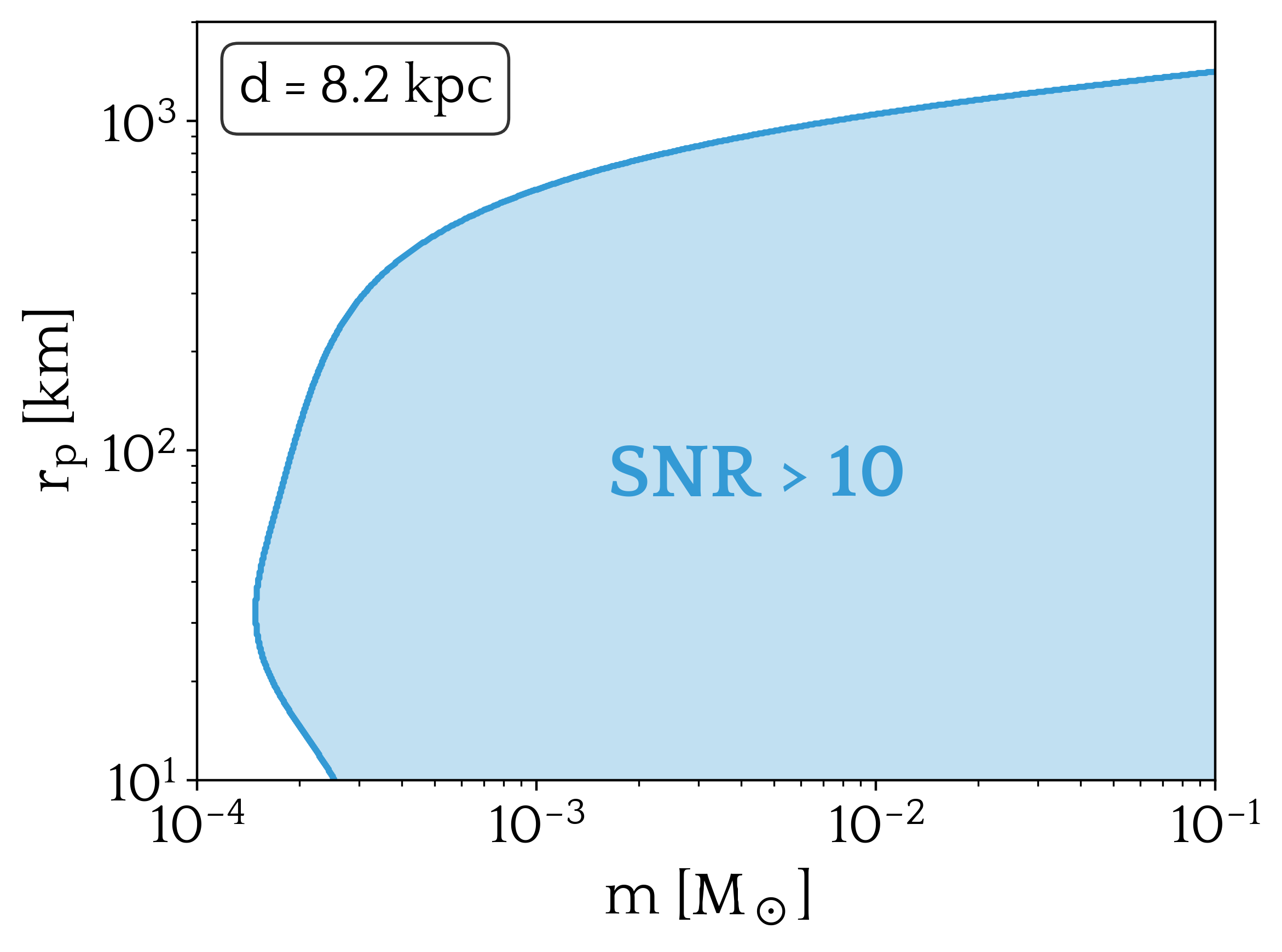}
    \caption{Periastron distances of PBHs orbiting neutron stars in elliptic or hyperbolic orbits with eccentricity close to $1$, for which an $\mathrm{SNR}>10$ will be generated in LVK O4, as a function of the PBH mass. The distance of the PBH-NS systems with respect to the Earth is $d=8.2$ kpc.}
    \label{fig:critical_SNR}
\end{figure}

Let us assume that a GW signal can be confidently identified when $\mathrm{SNR}>10$, see e.g. Ref~\cite{Bartolo:2016ami}. We can then plot, for a fixed distance $d=8.2~\text{kpc}$, the periastron distances that yield $\mathrm{SNR}>10$ as a function of the PBH mass. The result is displayed in Fig.~\ref{fig:critical_SNR}. Note that this calculation is valid for both bound and unbound PBH orbits. 

It is clear from Fig.~\ref{fig:critical_SNR} that GWs will only be detected if PBHs pass at a distance of less than $\lesssim1000$ km from neutron stars. We also see that GWs from PBH-NS encounters in the Galactic center can only be detected on Earth for PBH masses $m\gtrsim 10^{-4}M_\odot$. This mass threshold is one order of magnitude higher than that found in Ref.~\cite{Zou__2022}, which studied PBHs inspiralling {\em inside} neutron stars at the Galactic center. The difference is mainly due to their less conservative estimate, as they consider only the ratio between the GW and noise amplitude spectral densities. By requiring a minimum $\mathrm{SNR} > 10$, we impose a more stringent constraint on the signal.

\section{PBHs bound to neutron stars}
\label{sec:boundPBHs}

We now turn to determining how often bound PBH-NS pairs form. One possible formation channel is capture during star formation \cite{Capela_2012,Capela_2014,Oncins_2022,Esser_2022,Tinyakov_2024}. In this mechanism, close PBH-NS pairs are created by the gravitational pull of the contracting gas cloud during star formation. While this process is efficient in the asteroid-mass range \cite{Capela_2012,Capela_2014}, it does not operate for PBHs with masses $\gtrsim 10^{-4}M_\odot$: since neutron stars form from massive main-sequence stars, PBHs must be captured during the formation of the latter. For heavy PBHs, however, accretion onto the PBH is so efficient that a star that captures such a PBH will be converted into a black hole within a few days, without ever forming a neutron star.

Another way to form PBH-NS bound states is the direct capture mechanism \cite{Capela_2013}: when an unbound PBH passes near or through a neutron star, it can lose energy via processes such as gravitational radiation or dynamical friction, potentially acquiring a negative total energy and settling on a bound orbit. The rate at which PBHs with fixed (specific) asymptotic energy $E_\infty>0$ and angular momentum $L_\infty$ enter a sphere of arbitrary radius is given by 
\begin{equation}
\frac{dN}{dE_\infty dL_\infty^2dt}=4\pi^2n_\text{PBH}\left(\frac{3}{2\pi\bar{v}^2}\right)^{3\over 2}\exp\left(\frac{-3E_\infty}{\bar{v}^2}\right),
    \label{eq:rate_Press}
\end{equation}
see Eq.~(2.7) in \cite{Press_1985}\footnote{Note that a factor 2 was missing in the original calculation, as pointed out in \cite{Gould_1987}.}.
We denote $\Delta E$ (defined positive and larger than $E_\infty$) and $\Delta L$ as the total energy and angular momentum losses of the PBH during a close encounter. As we will argue below, for very eccentric orbits relevant in what follows, the relative energy change dominates over the angular momentum change, $\Delta E/E_\infty\gg\Delta L/L_\infty$. Moreover, for such orbits the energy loss $\Delta E$ does not depend on the PBH asymptotic energy $E_\infty$, but only on the squared angular momentum $L_\infty^2$. It is then possible to recast the rate \eqref{eq:rate_Press} as the rate at which PBHs are injected onto bound orbits around the NS with fixed periastron $r_p$ and apastron $r_a$. Due to the above property of $\Delta E$, the Jacobian for the change of variables $(E_\infty,L_\infty^2)\rightarrow(r_p,r_a)$ is independent of the energy loss, allowing one to write
\begin{align}
\begin{split}
        \frac{dN}{dr_pdr_adt}=&\frac{n_\text{PBH}}{\bar{v}^3}\frac{3\sqrt{6\pi}}{2}R_g^2\\&\times\frac{r_a-r_p}{(r_a+r_p)^3}\exp\left(\frac{-3E_\infty}{\bar{v}^2}\right)
        \label{eq:rate_direct}
\end{split}
\end{align}
with $R_g=2GM$. Considering for simplicity the case where all the neutron stars in the Milky Way lie in its center, with $N_\text{NS}=10^9$, the total injection rate of PBHs onto bound orbits around neutron stars is obtained by multiplying Eq.~\eqref{eq:rate_direct} by the number of NSs.

The above calculation assumes a Maxwellian distribution of relative velocities between PBHs and neutron stars, as motivated by the standard halo model for the DM distribution in the Milky Way, see e.g. \cite{Baxter_2021}. The actual distribution in the Galactic center, however, is subject to uncertainties and may deviate from this standard assumption \cite{lilie_2025}. Varying the velocity distribution within these uncertainties would only introduce corrections of $\mathcal{O}(1)$ to our results.

\section{PBH Energy loss and orbit evolution}
\label{sec:furtherevoPBH}
The same energy loss mechanisms responsible for the initial capture will also further extract energy from the PBH during its subsequent passages close to the neutron star, and dictate the evolution of the PBH orbit. As mentioned above, we consider two mechanisms for energy loss: when the PBH does not cross the NS, it looses energy through gravitational wave emission. When it crosses the NS, the dominant energy loss arises from the dynamical friction exerted by the NS material.

In the following, we focus on highly eccentric trajectories with $r_a \gg r_p$, which dominate the event rate. In this regime, $E = -R_g/(2r_a)$ and $L^2 = R_g r_p$. Moreover, since dissipative forces act primarily near periastron, the energy loss $\Delta E$ depends only on $r_p$ (equivalently, on $L$). One can also show that in this case, $\Delta E/E \sim  (r_a/r_p)  \Delta L/L \gg\Delta L/L$ regardless of the energy loss mechanism, in agreement with the assumptions made in Sec.~\ref{sec:boundPBHs}.

\subsection{Gravitational waves}
The energy loss rate via gravitational radiation in a system of two gravitating bodies was originally studied in \cite{Peters_1964}. The (specific) energy loss per period is obtained by multiplying their mean loss rate by the Keplerian period of an orbit and taking the large eccentricity limit, 
\begin{align}
\begin{split}
    &\Delta E_\text{GW}=\frac{85\pi}{192}\frac{R_g^6r_g}{c^5L^7}=\frac{85\pi}{192}\frac{1}{c^5}\frac{R_g^{5/2}r_g}{r_p^{7/2}}\\
    &\sim 10^{-12}\left(\frac{M}{M_\odot}\right)^{5/2}\left(\frac{m}{10^{-3}M_\odot}\right)\left(\frac{1000\text{ km}}{r_p}\right)^{7/2},
\label{eq:Eloss_GW}
\end{split}
\end{align}
where we defined $r_g=2Gm$. Most of the energy is lost during the swift passage of the PBH near the NS, and this formula therefore represents the energy dissipated per single close encounter. By analytic continuation, it is clear that the same expression also applies to unbound orbits, i.e. when the PBH approaches from infinity, as in the direct capture scenario.

\subsection{Dynamical friction}
When the PBH crosses the neutron star, the dense NS material applies a dynamical friction force onto the PBH, with magnitude given by the standard Chandrasekhar dynamical friction formula \cite{Chandra_1,BinneyTremaine}. For simplicity, we assume that the NS has a uniform density, that the length of the PBH trajectory inside the star is $\sim 2R$ (as would be the case for zero PBH angular momentum), and that its velocity during the crossing equals the escape velocity from the NS surface. Under these assumptions, we obtain
\begin{align}
\begin{split}
    \Delta E_\text{DF}&=\frac{3r_g}{2R}\ln\Lambda\\&\sim 5\times10^{-3}\left(\frac{m}{10^{-3}M_\odot}\right)\left(\frac{10\text{ km}}{R}\right),
\label{eq:Eloss_DF}
\end{split}
\end{align}
where the numerical estimate assumes $\ln\Lambda=\ln(M/m)\sim 10$. It is clear from comparing Eqs.~\eqref{eq:Eloss_GW} and \eqref{eq:Eloss_DF} that dynamical friction is the dominant energy loss mechanism when a PBH crosses a neutron star, except in rare cases when it only grazes the surface. 

\subsection{PBH inspiral time}

As explained in the previous subsection, in the high-eccentricity limit the energy lost by a PBH 
during each passage close to the NS depends only on $r_p$. The approximate conservation 
of $L$ relative to $E$ implies that $r_p$ remains (approximately) constant, while $r_a$ shrinks 
after each encounter. Consequently, the energy lost per passage is constant, and after $n$ passages the total energy is $E_n=E_0-n\Delta E$. Using Kepler's law, the orbital period after $n$ passages is then
\begin{equation}
    T_n=\frac{\pi}{\sqrt{R_g}}\left(\frac{R_g/2}{n\Delta E-E_0}\right)^{3/2},
    \label{eq:period_n}
\end{equation}
and the total inspiral time of a PBH is obtained by summing over the periods as \cite{Holst_2025,Tinyakov_2024}
\begin{equation}
    t_\text{ins}=\sum_{n=0}^\infty T_n = \frac{\pi}{2\sqrt{2}} \frac{R_g}{\Delta E^ {3/2}}\zeta\left(\frac{3}{2},\frac{-E_0}{\Delta E}\right)
\label{eq:merger_time}
\end{equation}
where $\zeta$ is the Hurwitz zeta function with asymptotics
\begin{equation}
    \zeta({3/ 2},x) \simeq \left\{
    \begin{array}{cc}
      x^{-3/2}    &  \text{for $x\ll 1$}  \\ 
      2x^{-1/2}   & \text{for $x\gg 1$}.
    \end{array}
    \right.
\end{equation}
For a PBH with initial apastron $r_a=350$ AU, the inspiral time is $\sim10^5$ yr for $\Delta E=10^{-12}$. However, when $\Delta E\gtrsim10^{-9}$, the inspiral time effectively becomes comparable to the orbital period $T_0$ corresponding to the initial apastron. In this regime, the energy loss is so large that the PBH orbital radius is drastically reduced by each passage, and only a few passages are required for it to merge with the neutron star.

\subsection{PBH deviations}
\label{sec:deviation}

To complete the calculation, it remains to estimate the maximum initial size of the orbit 
$r_a$. The value of $r_a$ is limited by gravitational perturbations from nearby stars which may deviate the PBH falling onto an NS and put it on a trajectory with large periastron, for which the energy losses are negligible. Given the high relative velocities of stars in the Galactic center compared to the typical orbital velocities of PBHs around NSs, the impulse approximation applies \cite{BinneyTremaine}. Over time, the effects of different perturbers add incoherently. Using Eq.~(17) of \cite{Esser_2025}, the mean square change of the angular momentum from external perturbers of mass $M_\ast$ during a single PBH orbit of period $T$ can be estimated as follows:
\begin{equation}
    \Delta L^2\sim \frac{2\pi^3G^2M_\ast^2 n_\ast r_a^2T}{\sigma_\ast}\ln\left(\frac{b_\text{max}}{b_\text{min}}\right).
\label{eq:delta_L2}
\end{equation}
Here $n_\ast$ is the local density of stars, $\sigma_\ast$ their relative velocity dispersion, $b_{\text{min}} = (\pi n_\ast \sigma_\ast T)^{-1/2}$  the impact parameter below which there is, on average, less than one perturber passage per PBH orbit, and $b_{\text{max}} = \sigma_\ast T$ the impact parameter above which the interaction time of perturbers with the PBH becomes larger than the orbital period, thereby breaking the impulse approximation. Note that $T$ depends on $r_a$ through Kepler's law.

Making use of the one-to-one relation between angular momentum and periastron distance in the highly eccentric regime, Eq.~\eqref{eq:delta_L2} can be translated into the corresponding increase of the periastron induced by stellar flybys, $\Delta r_p=\Delta L^2/R_g$. Requiring that this increase is small, i.e. $\Delta r_p<r_p$, leads to a condition on the maximum allowed apastron distance $r_a$ which has to be smaller than some critical distance $r_c$. For an environment typical of the inner kiloparsec of the Milky Way, with velocity dispersion $\sim 200~\mathrm{km\,s^{-1}}$ and stellar density $\sim 10~\mathrm{pc^{-3}}$ \cite{McMillan_2017}, we find $r_c \simeq 350~\mathrm{AU}$. In general, this condition also depends on $r_p$, but we find that this dependence is mild, with $r_c$ ranging from $r_c = 300~\mathrm{AU}$ (for $r_p=0$) to $r_c = 440~\mathrm{AU}$ (for $r_p=1000~\mathrm{km}$). For simplicity, we therefore fix $r_c = 350~\mathrm{AU}$, independently of the value of $r_p$. Furthermore, changing the stellar density by one order of magnitude modifies $r_c$ only by a factor of $\sim 2$, limiting the impact of uncertainties in this parameter. 

While the above estimate focuses on fast perturbers, one may instead compute the effect of slow perturbers on the PBH orbit. Using  Eq.~(6) of \cite{Esser_2022}, who employed the opposite approximation of static perturbers, one finds a value of $r_c \simeq 1100~\mathrm{AU}$ --  larger but within a factor of $\sim 3$ of our result, thereby indicating that the effect of these slow perturbers can be safely neglected. Moreover, typical wide binaries in the Solar neighborhood can have semi-major axes of up to several tens of thousands of AU \cite{Raghavan_2010}, but systems beyond that scale are fully disrupted by their environment. Our condition -- more stringent, since it does not require unbinding of the PBH but merely that the perturber-induced change in periastron remain small -- is also consistent with this constraint.

\section{Probability of signal detection}
\label{sec:probasignal}
We now have all the ingredients needed to compute the probability of detecting a GW signal from PBH-NS bound orbits in the Galactic center in a given observational period $t_\text{obs}$. We have seen above that only orbits with periastron $r_p$ and apastron $r_a$ in a certain range are relevant. Most of these orbits have a lifetime shorter than the age of the Milky Way, so we will use the steady-state approximation. In this regime, the total Galactic population of orbits that {\em have been created} with given $(r_p, r_a)$ and are currently still evolving toward final merger is determined by the product of the rate at which PBHs populate bound orbits (Eq.~\eqref{eq:rate_direct}) and the lifetime (total inspiral time) $t_\text{ins} \equiv t_\text{ins}(r_p,r_a)$ (Eq.~\eqref{eq:merger_time}) of PBHs on these orbits.

At a given moment, all these PBHs are randomly distributed with respect to their remaining  inspiral time, and occasionally produce a GW signal when passing through periastron. The probability to detect at least one of these signals within a time window $t_\text{obs}$ is given by the number of PBHs multiplied by some probability $P_S$. The latter quantity is the probability that at least one periastron passage will occur within $t_\text{obs}$ for a randomly picked PBH (that is, having random time remaining until the merger) with given original orbit parameters $(r_p, r_a)$. Importantly, complications related to the correlated nature of signals are hidden in $P_S$, while different orbits with different creation parameters $(r_p, r_a)$ are uncorrelated. One may therefore combine their contributions in a standard way using Poisson statistics and write the total probability to detect at least one signal from the {\em whole} PBH population as 
\begin{equation}
\label{eq:poisson}    
P = 1 - \exp \left( - \bar{N} \right), 
\end{equation}
where $\bar{N}$ is the mean number of PBHs that will produce at least one signal in the period $t_\text{obs}$. This number is obtained by integration over the whole relevant range of parameters $(r_p, r_a)$. We will see below that it is very small, so $P=\bar{N}$ to a very good approximation. 

Different energy loss channels -- dynamical friction and GW emission -- dominate in distinct regions of parameter space and therefore contribute additively, such that $\bar N = \bar{N}_\text{DF} + \bar{N}_\text{GW}$. We now turn to calculating these two contributions separately.

\subsection{Dynamical friction: direct mergers}
When PBHs cross neutron stars and lose energy via dynamical friction, we see from Eqs.~\eqref{eq:Eloss_DF} and \eqref{eq:merger_time} that the energy loss is so strong that they quickly merge onto the NS, without orbiting around it for an extended period of time. Each PBH therefore emits a single signal, corresponding to its merger with the NS. In this situation, $\bar N_\text{DF}$ is simply the number of PBHs that will merge in the time window $t_\text{obs}$,
\begin{equation}
\bar{N}_\text{DF}=t_\text{obs}\times\int\limits_{r_p,r_a}\frac{dN}{dr_pdr_adt}  \,dr_p\,dr_a,
\label{eq:N_DF}
\end{equation}
where we integrate over all periastrons and apastrons that eventually lead to merger, i.e. with $r_p<R$ and $r_a<r_c$. For the planetary mass PBHs considered here, the latter radius is set by the deviation condition (see Sec.~\ref{sec:deviation}) and is $r_c \sim 350\text{ AU}$. Using the injection rate from Eq.~\eqref{eq:rate_direct}, this integral can be evaluated analytically, yielding
\begin{align}  
\begin{split}
\bar{N}_\text{DF}=t_\text{obs}&\frac{n_\text{PBH}}{\bar{v}}N_\text{NS}\sqrt{6\pi}R_gR\\&\times\left(1-\exp\left[-\frac{9r_g\ln\Lambda}{2\bar{v}^2R}\right]\right),
\label{eq:N_DC_SF}
\end{split}
\end{align}
where we have also imposed $r_a>MR/(3m\ln\Lambda)$. This condition follows from requiring that the dynamical friction energy loss in Eq.~\eqref{eq:Eloss_DF} is such that $E_\infty$ in Eq.~\eqref{eq:rate_direct} is positive. In this case, the integral is in fact dominated by small apastron distances close to this lower bound and is independent of the upper bound $r_c$. This result is consistent with \cite{Capela_2014,Tinyakov_2024}. For the typical parameters considered here, the term in the exponential is large, and the exponential can be set to zero.

\subsection{Gravitational waves: correlated signals}
When PBHs do not cross NSs, GW emission drives the evolution, and the inspiral time $t_\text{ins}$ is typically way larger than the observation time $t_\text{obs}$. In this case, one can compute the probability $P_S$ as follows. When one randomly places the observation window of duration $t_\text{obs}$ within the inspiral time period, two situations are possible. If the PBH is in the early phase of its inspiral, consecutive periastron passages are separated by a time interval longer than $t_\text{obs}$, so that each passage contributes $t_\text{obs}/t_\text{ins}$ to $P_S$. Alternatively, in the final phase of the PBH inspiral periastron passages occur more frequently than one per $t_\text{obs}$, in which case the contribution to 
$P_S$ is the total duration of this part of the inspiral, divided by $t_\text{ins}$. In summary,
\begin{equation}
    P_\text{S}=\frac{1}{t_\text{ins}}\left[n_ct_\text{obs}+\frac{\pi}{2\sqrt{2}} \frac{R_g}{\Delta E^ {3/2}}\zeta\left(\frac{3}{2},n_c-\frac{E_0}{\Delta E}\right)\right]
\end{equation}
where $n_c$ is the number of passages after which the orbital period of the PBH becomes shorter than $t_\text{obs}$; it is obtained by solving $T_n=t_\text{obs}$ using Eq.~\eqref{eq:period_n}. The first contribution corresponds to PBHs in the early phase of the inspiral, while the second arises from PBHs in the late phase.

The mean number of PBHs currently orbiting neutron stars is given by the product of the injection rate and the duration of the inspiral $t_\text{ins}$. The mean number of PBHs that will emit at least one signal in the next $t_\text{obs}$ years is the product of the mean number of PBHs and $P_S$, that is
\begin{equation}
\bar{N}_\text{GW}=\int\limits_{r_p,r_a}\frac{dN}{dr_pdr_adt}\, t_\text{ins}\, P_\text{S}\,dr_p\,dr_a,
\label{eq:mean_signal_number_GW}
\end{equation}
which is integrated over the possible initial apastrons and periastrons. Note that, in some regions of the parameter space where $\Delta E$ is too large, it may happen that $n_c\rightarrow 0$, in which case we have $P_S\rightarrow t_\text{obs}/t_\text{ins}$, such that Eq.~\eqref{eq:N_DF} is recovered from Eq.~\eqref{eq:mean_signal_number_GW}.

For the direct capture mechanism (Eq.~\eqref{eq:rate_direct}), we compute the integral \eqref{eq:mean_signal_number_GW} numerically. Note that every factor in the integrand is a function of $r_p$ and $r_a$. The bounds of integration are set by requiring $r_a<r_c=350\text{ AU}$ and $r_p$ such that SNR $>10$. 

Because we neglected in our calculations the evolution of $r_p$ during PBH inspirals, PBHs with initial periastrons exceeding $1000\text{ km}$ do not contribute into the GW signal rate (cf. Fig.~\ref{fig:critical_SNR}). We have verified the validity of this approximation by estimating the contribution of bound PBHs which start with $r_p>1000\text{ km}$ and then enter the relevant range $r_p<1000\text{ km}$ due to gradual decrease of $r_p$ during inspiral. We found that this contribution is negligible compared to the total signal rate.

\section{Results and conclusions}
\label{sec:conclusion}
\begin{figure}[t!]
    \centering
    \includegraphics[width=\linewidth]{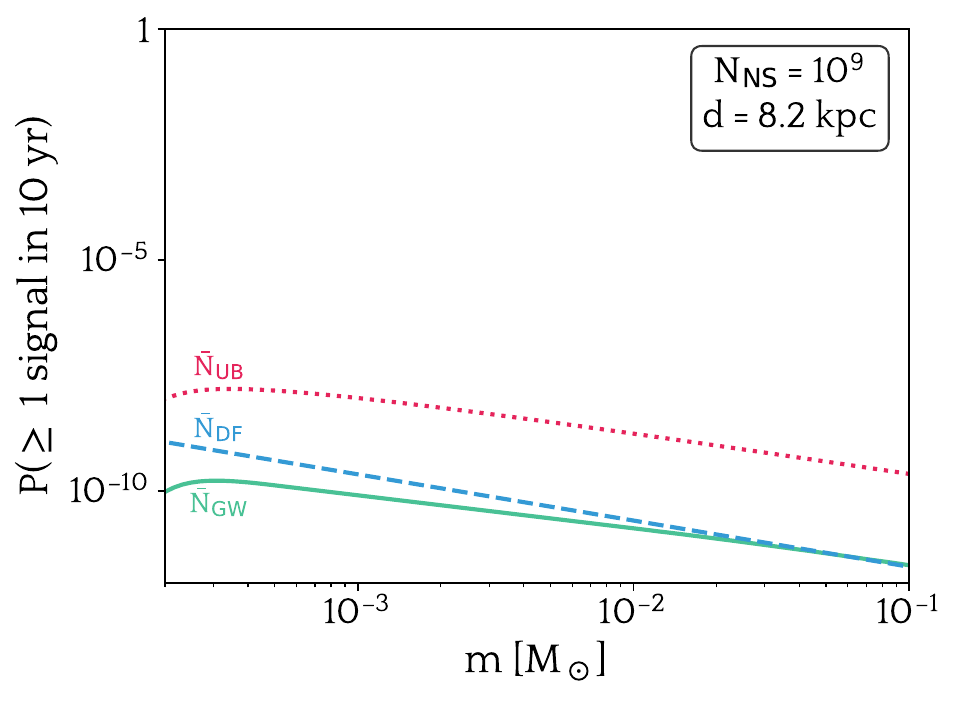}
    \caption{Probability of detecting GW signals from PBH-NS pairs in the Galactic Center over 10 years of observation with LVK O4 sensitivity, using fiducial values $\rho_\text{DM}=10\text{ GeV}/\text{cm}^3$ and $\bar{v}=\sqrt{2}\times200\text{km}/\text{s}$ for the Galactic properties. The most likely signals to be detected correspond to PBHs on unbound orbits passing close to neutron stars (dotted red). Next are PBHs bound to neutron stars and losing energy via dynamical friction (dashed blue), followed by those losing energy via gravitational wave emission (plain green).}
    \label{fig:detec_proba}
\end{figure}

In Fig.~\ref{fig:detec_proba}, we plot the probability of a signal
detection in a 10-yr observational period, assuming the sensitivity of the  LVK O4 run.  Probabilities are shown separately for two energy loss channels as discussed in Sec.~\ref{sec:probasignal}. For comparison, we also include the signal probability from unbound PBH-NS encounters (dotted red curve), for which $\bar{N}_{\text{UB}} = \Gamma\, t_{\text{obs}}$. Here $\Gamma$ is given by Eq.~\eqref{eq:rate_unbound}, with the detection radius $r_\text{det}$ for a given mass $m$ obtained from Sec.~\ref{sec:strain} by requiring SNR $>10$. We use fiducial parameters $\rho_\text{DM}=10\text{ GeV}/\text{cm}^3$ and $\bar{v}=\sqrt{2}\times200\,\text{km}/\text{s}$ for the Galactic center. 

Because the signal probability in the case where the dynamical friction drives the energy loss (dashed blue curve) is given by the merger probability, it decreases with the PBH number density as $\propto m^{-1}$ (see Eq.~\eqref{eq:N_DC_SF}). However, the mass dependence of $\bar N_\text{GW}$ (plain green curve) is less trivial because of the nonlinear dependence of the SNR on the mass (see e.g. Fig.~\ref{fig:critical_SNR}).

Below $ m \sim10^{-4} M_\odot$, PBHs are too light to produce detectable GWs in the Milky Way, unless they would be located relatively close ($\lesssim1$ kpc) to Earth; however, this is unlikely given the small number of neutron stars within such a nearby volume. PBH masses below this threshold are therefore not displayed on the figure.

We find that signals from bound PBH-NS pairs are subdominant relative to the probability of detecting an unbound encounter. Although repeated periastron passages of bound PBHs enhance their detection probability, the  number of bound systems is too low, and hence unbound PBH-NS encounters 
dominate the probability. Such grazing encounters are therefore the most promising channel for PBH detection compared to other scenarios, even though, given the current sensitivity of GW detectors, the probability that a detectable event will occur in the next ten years is very small, $P \sim 10^{-8}$. This should be taken into account in future theoretical studies of possible GW signals from PBHs.

One may think of several scenarios in which the small detection probability we found could be at least partially compensated. While we have assumed that DM is uniformly distributed in the Galactic center, PBHs may in fact be significantly clustered in dense pockets around which stars may preferentially form. One can estimate the mass and size of such clusters \cite{Carr:2023tpt}, from which the corresponding PBH number density can be inferred. We find that, for the typical PBH masses we consider, their density within these clusters could be $\sim 100$ times higher as compared to our estimate based on $\rho_\text{DM}=10 \text{ GeV}/\text{cm}^3$ in the Galactic center. If most of the neutron stars reside within such PBH clusters, the resulting signal rate could be enhanced by the same factor.

Another possibility discussed in the literature is that the supermassive black hole at the center of our Galaxy may be surrounded by a dark matter spike \cite{Caiozzo_2024, Gondolo_1999}, as predicted if this black hole underwent adiabatic growth following the formation of the central region of the DM halo. In some extreme scenarios, the spike in the inner parsec of the Milky Way could reach dark matter densities of $\rho_\text{DM} \gtrsim 10^{6}\text{ GeV}/\text{cm}^3$, with a steep power-law profile characterized by an index $\gamma \sim 2 - 2.25$. Given the stellar density close to $\sim10^6\text{ pc}^{-3}$ in the inner parsec of the Milky Way \cite{Schodel_2007}, this region may contain up to $\sim 10^4$ neutron stars. Even by combining these optimistic estimates with a sensitivity improvement of roughly one order of magnitude expected from the upcoming Einstein Telescope~\cite{ET:2019dnz,ET:2025xjr}, the detection probability of planetary-mass primordial black holes would still remain below $\mathcal{O}(1)$ under these assumptions.

Given the low detection probabilities found here, a more viable way to use neutron stars as probes of primordial black holes is probably to rely on integrated effects on the neutron star population, i.e. indirect consequences of PBH-NS interactions that can be measured \textit{a posteriori} and whose effects accumulate over time. One example -- although relevant in a different PBH mass range -- is the PBH-induced destruction of the pulsar population in the Galactic center, which would suppress their observed signals \cite{Caiozzo_2024}. Another possibility is to search for potential r-process enhancement in stars due to the destruction of neighboring neutron stars by PBHs \cite{Fuller_2017}. Given the significant astrophysical uncertainties associated with these observables, however, they currently lack the robustness required to place firm constraints on PBHs.

\acknowledgements{NE is a FRIA grantee of the Fonds de la Recherche Scientifique-FNRS and a member of BLU-ULB. JGB acknowledges support of the Spanish Agencia Estatal de Investigaci\'on through the Research Project PID2024-159420NB-C43 [MICINN-FEDER], and the grant ``IFT Centro de Excelencia Severo Ochoa CEX2020-001007-S". 
PT is supported in part by the Institut Interuniversitaire des Sciences Nucl\'eaires (IISN) Grant No. 4.4503.15. \\[1mm]
Preprints: IFT-UAM/CSIC-25-169, ULB-TH/26-05.}

\bibliographystyle{apsrev4-2} 

\bibliography{bibli}

\end{document}